\documentclass[twocolumn]{emulateapj}

\usepackage{natbib}
\usepackage{amssymb}
\usepackage{graphicx}
\newcommand{\src}{GS 1826$-$238}
\newcommand{\rxte}{\it RXTE}

\begin{document}
\title{Deviations from the Flux-Recurrence Time Relationship in GS 1826--238:\newline
Potential Transient Spectral Changes}
\author{
Thomas W. J. Thompson\altaffilmark{1},
Duncan K. Galloway\altaffilmark{2},
Richard E. Rothschild\altaffilmark{1},
Lee Homer\altaffilmark{3}}
\altaffiltext{1}{Center for Astrophysics and Space Sciences, University of California, San Diego, La Jolla, CA 92093}
\altaffiltext{2}{School of Physics \& School of Mathematical Sciences, Monash University, Australia; Monash Fellow}
\altaffiltext{3}{Abbey College, Cambridge, UK}

\begin{abstract}
The low-mass X-ray binary \src~is presently unique for its consistently regular bursting behavior. In previous {\it Rossi X-Ray Timing Explorer} ({\it RXTE}) measurements between 1997 November and 2002 July, this source exhibited (nearly) limit-cycle bursts with recurrence times that decreased proportionately as the persistent flux increased. Here we report additional measurements of the burst recurrence time by {\it RXTE}, {\it Chandra}, and {\it XMM-Newton}, as well as observations of optical bursts. On a few occasions we measured burst recurrence times which deviated significantly from the earlier flux-recurrence time relationship, and most of these bursts occurred earlier than would be predicted based on the X-ray flux level. The epochs with early bursts were also accompanied by unusual broadband timing signatures, with the entire power spectrum shifting to higher frequencies. Concurrent {\it XMM-Newton} observations during one of these occasions, in 2003 April, indicate that an additional soft component may be present in the spectrum containing enough flux ($\approx$30\% of the total) to account for the burst recurrence time discrepancy. A self-consistent interpretation for the increase in soft flux and accompanying timing changes during 2003 April is that accretion disk extends down to smaller radial distances from the source than during the other observing epochs. The {\it RXTE} observations since 2003 April show that the spectral and timing properties have nearly returned to the previously established level. 
\end{abstract}

\section{Introduction}
Type I X-ray bursts result from unstable thermonuclear ignition of
accreted material on the surfaces of weakly magnetic ($B < 10^{10}$ G)
neutron stars (for a recent review, see \cite{sb06}). Freshly accreted
hydrogen and helium on the neutron star surface is hydrostatically
compressed by new material at a rate $\mbox{\.{\em m}} \sim 10^{4}$ g
cm$^{-2}$ s$^{-1}$. In systems exhibiting bursts, the temperature and
pressure at the base of the accreted layer slowly increase until the
nuclear energy generation rate of the $3\alpha$-reaction becomes more
sensitive to temperature perturbations than the radiative cooling. At this
point the resulting thermonuclear instability leads to runaway burning of
some or all of the matter that has been deposited since the previous
burst.
Typically, hours to days are required to accrete enough material to trigger the instability. 

There are close to 100 type I bursters that are known in the
Galaxy\footnote{For an updated list of Galactic type I X-ray bursters, see
http://www.sron.nl/$\sim$jeanz/bursterlist.html.}, and the vast majority
are ``atoll'' type low-mass X-ray binaries (LMXBs) with luminosities above
about $10^{36}$ erg s$^{-1}$ \citep{hv89}. Although the basic physics of
type I X-ray bursts is understood, detailed comparisons between
observations and theoretical models have had mixed success (e.g., Woosley
et al. 2004). The most successful comparison has been with the  ``clocked
burster'' \src~(also known as Ginga 1826--238), whose bursts are
consistently quasi-periodic \citep{uber99,cocchi01,corn03}. Studies of the
bursts from GS 1826--238 began during the {\it BeppoSAX} mission, with the
most detailed study of burst recurrence times and energetics coming from
\cite{g04}, who analyzed 24 bursts from the {\it Rossi X-Ray Timing
Explorer} ($\rxte$) detected between 1997 November and 2002 July. They
found that the recurrence time decreased from $5.74 \pm 0.13$~hr to
$3.56\pm0.03$~hr while the persistent (between burst) flux level increased
by 66\%, so that the recurrence time decreased almost precisely as $1/F_{x}$. Assuming $F_{x} \propto \mbox{\.{\em M}}$ implies that the accumulated mass required for the instability to occur is approximately the same even as the accretion rate changes.  The long burst durations ($\sim$100 s) and the low value of $\alpha$ ($\approx$40) -- the ratio of the integrated bolometric persistent flux between bursts to the total bolometric burst fluence -- both suggest that hydrogen constitutes a large portion of the fuel for these bursts \citep{bildsten00}. 
Further support for the H-rich fuel scenario comes from comparisons of the
observed light curves with those predicted by time-dependent models, which
also imply solar H and CNO composition in the material accreted by \src\
(Heger et al. 2008).

Additional $\rxte$~observations have been made as part of a monitoring campaign from 2003 through 2007. These data reveal that the previously monotonic relation between burst recurrence time and persistent X-ray flux no longer fully describes the source behavior. Here we present an analysis of the full data set with the objective of more fully characterizing the complex burst behavior. We have also analyzed observations from the {\em Chandra X-ray Observatory} in 2002 July, and {\em XMM-Newton} in 2003 April, both of which occurred simultaneously with $\rxte$, in order to also study the low-energy spectrum. In addition, we used optical observations made with the UCT-CCD fast photometer at the South African Astronomical Observatory (SAAO) in 1998 June to determine the bursting behavior during gaps in the $\rxte$~observations. In \S~2 we describe these observations and the details of our data analysis. We present our spectral and timing results in \S~3. Finally, we discuss the implications of the results in \S~4.

\section{Observations and Analysis}
\subsection{RXTE}
In this paper, we utilized all $\rxte$~observations of \src, which has been observed at least once per year since 1997, except during 2001. Overall, there are more than 600 ks of good exposure time. We analyzed data from the Proportional Counter Array (PCA; Jahoda et al. 2006) and the High Energy X-Ray Timing Experiment (HEXTE; Rothschild et al. 1998) instruments. The PCA has five identical co-aligned Proportional Counter Units (PCUs) sensitive to 2--60 keV photons. The HEXTE comprises two clusters, each containing four scintillation detectors sensitive to 15--250 keV photons. Both instruments have large effective areas ($\sim$6000 and 1400 cm$^{2}$, respectively) and microsecond timing resolution. 

\subsubsection{Recurrence Time Measurements}
Measurements of the burst recurrence times were made using PCA event mode data with 1 s bins, including all photon energies. The times of the bursts were defined to be when the flux exceeds 25\% of the peak flux of each of the bursts. Sometimes bursts took place when the satellite was not taking any data (e.g., during passages through the South Atlantic Anomaly), and recurrence times were inferred if bursts occurred near enough together to unambiguously know the number of intervening bursts. This method is acceptable because bursts from \src~have {\it never} been observed to occur at irregular intervals. If the time between two observed bursts is $\Delta t$, the recurrence time is estimated as $\Delta t/(n+1)$, where $n$ bursts are inferred in data gaps. Following Galloway et al. (2004), we adopted a fractional error of 2\% on the individual recurrence time measurements, and when bursts are inferred the error is scaled by $1/(n+1)^{1/2}$ because we assume $n+1$ burst intervals in total. Although {\it RXTE} observations took place during 1999 and 2005, the duration and spacing of these measurements did not allow for an unambiguous recurrence time measurement. Table 1 shows the observation dates containing recurrence time measurements, the number of bursts that were observed, and the average burst recurrence time during each epoch. The observations taking place simultaneously (or nearly so) with {\it RXTE} are discussed below.
\begin{deluxetable}{lccc} 
\footnotesize
\tablenum{1}
\tabletypesize{\scriptsize}
\tablecolumns{4}
\tablewidth{0pt}
\tablecaption{\sc{Observation Log of GS 1826--238}} 
\tablehead{
\colhead{Date} &
\colhead{} &
\colhead{} &
\colhead{Recurr. Time}\\
\colhead{(U.T.)} &
\colhead{Observatory\tablenotemark{a}} &
\colhead{Bursts\tablenotemark{b}} &
\colhead{(hr)} 
}
\startdata   
1997 Nov 5--6 & RXTE & 2 & 5.88\tablenotemark{c} \\
1998 Jun 7--12 & RXTE & 3 & 5.67 $\pm$ 0.04 \\
1998 Jun 23--25 & SAAO/RXTE & 6/1 & 5.13 $\pm$ 0.19 \\
2000 Jun 30--03\tablenotemark{d} & RXTE & 8 & 4.04 $\pm$ 0.06 \\
2000 Sep 24--27 & RXTE & 7 & 4.11 $\pm$ 0.06 \\
2002 Jul 29--30 & Chandra/RXTE & 6/4 & 3.54 $\pm$ 0.03 \\
2003 Apr 6--9 & XMM/RXTE & 16/7 & 3.18 $\pm$ 0.14 \\
2004 Jul 19--21 & RXTE & 7 & 3.47 $\pm$ 0.02 \\
2006 Aug 9--12 & RXTE & 11 & 3.36 $\pm$ 0.07 \\
2007 Mar 8--10 & RXTE & 9 & 3.53 $\pm$ 0.02 \\                               
\enddata
\tablecomments{Observations allowing unambiguous recurrence time measurements. Additional {\it RXTE} observations are utilized in our study of the power spectra (see \S~\ref{ta}).}
\tablenotetext{a}{During three periods, simultaneous observations took place with {\it RXTE}.}
\tablenotetext{b}{All bursts observed by {\it RXTE} were also observed with the other observatory.}
\tablenotetext{c}{Only one burst interval was measured during this epoch.}
\tablenotetext{d}{This observation ended in July.}
\end{deluxetable}

\subsubsection{Energy Spectra Analysis}
\label{esa}
The evolving spectrum during the bursts was modeled using an absorbed
blackbody, subtracting the pre-burst persistent emission as background. This approach is relatively standard for X-ray burst analysis, and we refer the reader to \cite{g04} for further details. In addition to the burst spectra, we studied the persistent emission by extracting color-color diagrams and photon energy spectra. In each case we excluded data from 500 s before to 1500 s after a burst, the latter chosen to minimize the residual blackbody flux seen during the decays of the bursts \citep{thompson05}. Colors were created by extracting light curves for each PCA channel and observation, reading in the channel energy boundaries from the observation's response matrix, and interpolating the counts spectrum to a standard grid. The soft and hard colors are defined as the counts ratio (3.5--6) keV/(2--3.5) keV and (9.7--16) keV/(6--9.7) keV, respectively, and were obtained by integrating over the interpolated grid. To correct for the long-term drift of the gain of each PCA channel, the colors were normalized to the colors of the Crab calculated with the closest observation available to each \src~observation.  

We extracted energy spectra from the persistent emission between X-ray bursts using the standard software for $\rxte$~data reduction (FTOOLS v.6.2). For the PCA, we used the ``Standard 2'' data, ignoring the first three channels and photon energies greater than 30 keV. HEXTE data below 25 keV were ignored, as were data above 100 keV due to poor statistics. The HEXTE channels were rebinned so that there were a minimum of 1000 counts per bin. To each observation (minus the times surrounding the bursts), we fit an absorbed double Comptonization model using XSPEC v.11.3. This model was found by \cite{thompson05} to fit the broadband spectra of \src~better than a single Comptonization model or empirical models like a cut-off power-law. For each measured $\rxte$~burst recurrence time, the persistent flux was obtained from the best-fit model\footnote{Some of the fluxes had to be corrected for a 0$\fdg$251 offset pointing in R.A., which decreased the efficiency of the PCA and HEXTE by a factor of $\sim$1.3.} during the time since the previous burst. Although the spectral model only applies to photons from 3--100 keV, we extrapolated the model from 0.1--3 keV and 100--200 keV to estimate the bolometric flux. 

\subsubsection{Timing Analysis}
\label{ta}
To relate the burst behavior to the broadband (persistent) timing
properties,
we also analyzed the rapid variability in the X-ray emission by producing a series of power density spectra (PSDs). For each {\it RXTE} observation, including six observations that did not allow for a recurrence time measurement, we used 128 s segments of PCA event mode data with $2^{-12}$ s bins (corresponding to a Nyquist frequency of 2048 Hz), and normalized after \cite{leahy83}. As with the energy spectra and colors, we excluded data 500 s before to 1500 s after each burst. Separate groups of power spectra close in time were merged if no systematic differences in PSDs were observed. In this manner we obtained 18 power spectra. On average, each PSD contained 36 ks of accumulated data (with the smallest being 8 ks and the largest being 83 ks). The contribution due to Poissonian statistics was estimated from the 1500--2048 Hz frequency band and removed (this also implicitly accounts for any decrease in power due to the PCA deadtime), and the spectra were converted to fractional rms squared. 

Qualitatively, each PSD can be characterized as having roughly equal power
per decade in frequency, which extends between a low and high frequency
break. To quantitatively describe each power spectrum, we fit a combination of 4--6 Lorentzians following, e.g., \cite{bpv02}, which overcomes the limitation of treating different power spectral components with intrinsically different models.  Two to three of the Lorentzians are zero-centered and fit the broadband noise, giving a broad peak in the power times frequency ($\nu P_{\nu}$) representation between $\nu_{b}$ at low frequencies and $\nu_{l}$ and $\nu_{u}$ at high frequencies \citep{bpv02}. There is also an additional component peaked at $\nu_{h}$ to cover the ``hump'' typically seen near 2--3 Hz (but much higher during 2003 April, as shown below) in \src. Additional narrow Lorentzians that are interpreted as quasi-periodic oscillations (QPOs; where the coherence factor $Q \equiv \nu_{0}/2 \Delta >2$) were added if necessary to achieve acceptable fits. Due to limited statistics at high frequencies, only the lower frequency Lorentzian components ($< 200$ Hz) were useful for comparison. The frequencies are characterized using  $\nu_{\rm max}$, the frequency at which the component contributes most of its variance per logarithmic frequency interval, which is equal to $\nu_{\rm max} = \sqrt{\nu_0^2+\Delta^2}$, where $\nu_0$ is the centroid frequency and $\Delta$ is the half-width at half-maximum of the Lorentzian. Errors on the fit parameters were determined using $\Delta \chi^2=2.71$ corresponding to the 90\% confidence interval. 

\subsection{Chandra}
\label{chandrasect}
The {\it RXTE}\/ observations on 2002 July 29--30 occurred simultaneously with {\em Chandra} \cite[also analyzed by][]{thompson05}, providing valuable
low-energy coverage. A 68 ks observation was made with the Advanced CCD Imaging Spectrometer (ACIS; Garmire et al. 2003), which is sensitive to photons from 0.3--10 keV. High-resolution spectra were obtained by having the High Energy Transmission Grating (HETG; Canizares et al. 2005) placed in the optical path. The HETG spectrometer is composed of the Medium-Energy Grating (MEG) and the High-Energy Grating (HEG). First-order MEG/HEG spectra were extracted with the {\em Chandra} Interactive Analysis of Observations (CIAO) v.3.4 software. Responses matrices were generated using calibration v.3.4.0. Spectra were rebinned with at least 1000 counts per spectral channel, giving statistically significant data for photon energies greater than 1 keV.

\subsection{XMM-Newton}
\label{xmmsect}
The {\it RXTE}\/ observations during 2003 April 6--9 occurred
simultaneously with an {\em XMM-Newton} (Aschenbach 2002) observation \cite[also analyzed
by][]{kong07}, lasting for 200 ks. Timing mode data was acquired from both the European Photon Imaging Camera (EPIC) and the Reflection Grating Spectrometer (RGS), although we only made use of data from the EPIC pn detector. The Optical Monitor was turned off for this observation. Data analysis was performed with the {\em XMM-Newton} Science Analysis System (SAS) v.7.1.0. Photon energy spectra were extracted for energies between 0.5 and 6 keV, and were rebinned with at least 1000 counts per spectral channel. Rather than using the ``canned'' response matrices, we generated one specific to our observation.

\subsection{Optical}
During 1998 June, a set of three 7 ks $\rxte$ observations were made at $\sim$1 day intervals, resulting in the detection of just one burst. At the same time, however, observations of a small ($50\times 33$ arcsec$^{2}$) region surrounding the optical counterpart of \src~were made using the UCT-CCD fast photometer \citep{odon95}, at the Cassegrain focus of the 1.9 m telescope at SAAO. On June 23 and 24, the observations lasted for 8 hours in an uninterrupted series of 5 s exposures; on June 25, the observation was 9 hours long using 2 s exposures. In each case, the coverage included (unambiguous) consecutive optical bursts from \src. We performed data reduction using the Image Reduction and Analysis Facility (IRAF), including photometry with the implementation of DAOPHOT II (Stetson 1987). Point spread function (PSF) fitting was employed in order to obtain the best possible photometry, since there was moderate crowding of the counterpart and variable seeing. The details of this procedure are given by \cite{homer98}. 

\begin{figure}
\centerline{\includegraphics[width=3.2in]{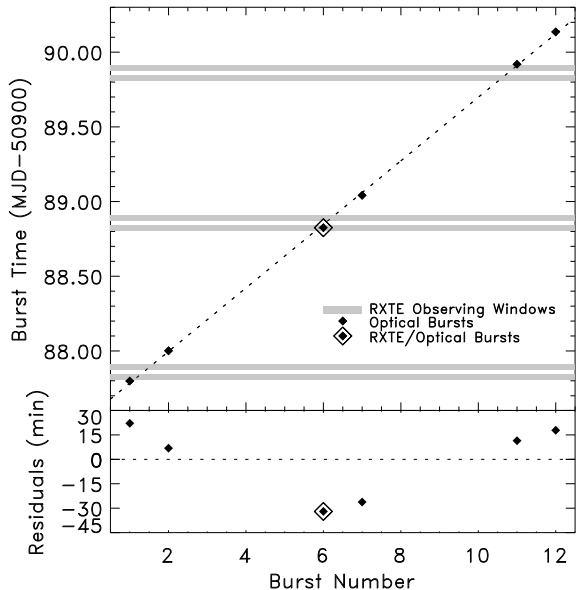}}
\caption{\footnotesize Burst arrival time plot during the optical observations in 1998 June. Intervening bursts (\#3--5 \& \#8--10) were inferred to take place with consistent separations, although since the inferred times were during daylight at SAAO they could not be observed. The shaded bars show the observing windows for the nearly concurrent {\em RXTE} observations. The {\em dashed} line assumes a constant burst recurrence time of 5.10 hr, although the burst behavior is more accurately described by a recurrence time of 4.92 hr for the first three observed bursts (1, 2, \& 6, assuming constant intervals between 2 \& 6), and 5.24 hr for the last four observed bursts (6, 7, 11, \& 12, assuming constant intervals between 7 \& 11), using burst 6 for each measurement. The apparent $V$-shaped residual plot is due to this change in recurrence time. If we assume the recurrence time changed discontinuously at burst 6, the residuals are only $\sim$2 min. \label{arrivalplot}}
\end{figure}

Figure \ref{arrivalplot} shows the arrival times of the bursts seen in the
optical and by {\it RXTE}. The bursts detected in the optical, coincident with (and immediately
following) the burst detected by {\it RXTE}\/ allowed an unambiguous
measurement of the persistent flux and recurrence time during these
observations. In addition, for the subsequent analysis, we associate the
other two recurrence time measurements made possible by the pairs of
bursts observed in the optical both before and after the burst detected by
{\it RXTE}, with the X-ray flux measured during that observation.

\section{Results} 
\begin{figure}
\centerline{\includegraphics[width=3.2in]{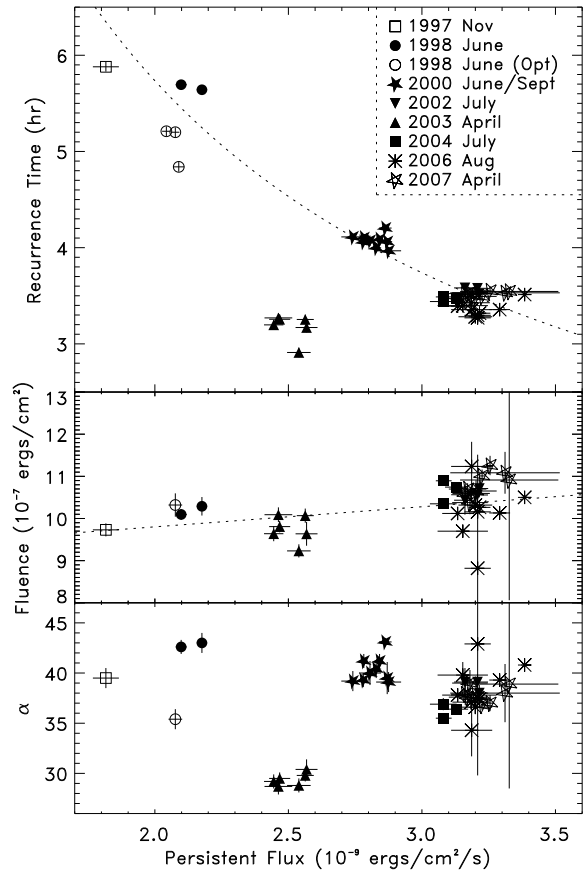}}
\caption{\footnotesize Variation of the burst recurrence time, burst fluence, and $\alpha$ ($\equiv \int_{0}^{\Delta t} F_{\rm p} dt/\int_{0}^{\Delta t} F_{\rm b} dt$) as a function of the absorbed persistent (between burst) flux in the range 0.1--200 keV (derived from extrapolating the model fit to {\it RXTE} data in the 3--100 keV range). Note that the unabsorbed fluxes are consistently $\sim$5\% larger. The {\em dotted} curve in the top panel shows the empirical fit to the data (excluding 2003 April), using $\Delta t \propto F_{x}^{-1.05}$ \citep{g04}; the middle panel contains a linear fit. The 1998 June (opt) measurements in the top panel use optically-measured recurrence times and X-ray fluxes from the nearest {\it RXTE}\/ observations (see Fig. \ref{arrivalplot}). The lower panels only have one corresponding data point because {\it RXTE}\/ measurements are only available for a single 1998 June (opt) burst.\label{fluxdt}}
\end{figure}
Within any individual observation, \src~exhibited quasi-periodic bursting
behavior, with the average persistent flux in between bursts varying by at
most a few percent. Figure \ref{fluxdt} shows the variation of the burst
recurrence time, burst fluence, and $\alpha$ as a function of the
(absorbed) persistent X-ray flux using {\it RXTE} measurements. Except for
the data from 2003 April, it is evident that the recurrence time of the
bursts has a rough $1/F_{\rm X}$ dependence \citep{g04}, as has been
observed previously, implying that typically, a similar amount of accreted
fuel is required to trigger each burst. 
The typical scatter on the recurrence times within each epoch is just 1\%.
However, the 2003 April measurements are significant outliers, with
recurrence times that are shorter than expected based on the persistent
flux.
During this epoch, the recurrence time was about 3.2 hr even though the
persistent flux was $\sim$10\% lower than in 2000 June/September, when the
recurrence time was about 4.1 hr. Other measurements also deviate from the
$1/F_{\rm X}$ trend, but to a lesser extent. For example, in 1998 June the
recurrence time was typically near 5.65 hr, but about twelve days later
optical burst intervals were 4.9 and 5.2 hr even though the persistent
X-ray flux (from $\rxte$ observations) had not substantially changed. Therefore, the occasional discrepancy between the burst recurrence time and the expected value (given the persistent flux level and their previously observed relationship) seems to be a general property of \src, and not something that (for example) only occurs near the highest persistent flux levels. More recent {\it RXTE}\/ data from 2004, 2006, and 2007, have shown that the persistent flux versus recurrence time relationship has at other times obeyed the previously observed monotonic behavior. 

\begin{figure}
\centerline{\includegraphics[width=3in]{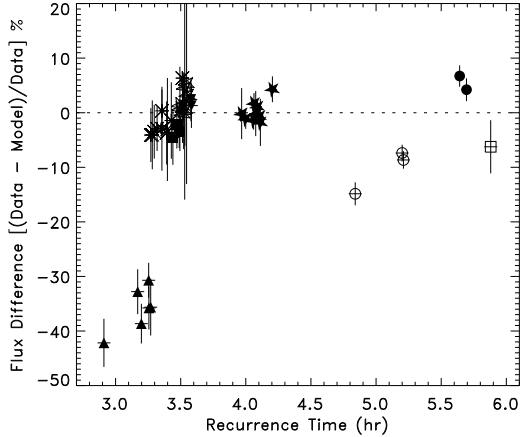}}
\caption{\footnotesize Fractional difference between the estimated bolometric flux and the ``expected'' flux derived from the empirical relationship between flux and burst recurrence time (``recurrence time flux'') as a function of the recurrence time. The symbols correspond to the same time periods as in Fig. \ref{fluxdt}.\label{resid}}
\end{figure}

Figure \ref{resid} shows the fractional difference between the estimated bolometric flux and the flux derived from the empirical relationship between flux and burst recurrence time (hereafter; referred to as the ``recurrence time flux''), as a function of burst recurrence time. Note that this figure can be equivalently considered as the residuals of the $x$-axis in Fig. \ref{fluxdt}, {\it top} panel. Assuming the empirical curve describes the actual flux-recurrence time relationship, Fig. \ref{resid} suggests that the 2003 April X-ray flux measured by {\it RXTE}\/ is underestimated by 30--40\%. 



The discrepant measurements suggest either that the proportionality between the persistent 0.1--200 keV flux (derived from extrapolating the model fit in the 3--100 keV range) and accretion rate was different than during other observations, or that the heating of the accreted layer (or some other factor) changes to alter the ignition depth at the same accretion rate. In order to discriminate between these two possibilities, we made a more detailed study of the burst characteristics, and of the persistent spectral and timing behavior of the source.

\subsection{The Energy Spectrum-Accretion Rate Relationship}
\label{es}
It is conventionally assumed that the accretion rate in LMXBs is
proportional to the X-ray flux, and since we measure the flux in an
instrumentally-determined passband, there is always the possibility that
spectral variations outside our passband will cause the proportionality to
change between observation epochs. Here we explore this possibility in
detail, both from the spectral shape (via the X-ray colors) within the
{\it RXTE}\/ passband (which is common to all the measurements), and in a
broader band possible thanks to several occasions of contemporaneous {\it
Chandra}\/ and {\it XMM-Newton}\/ observations.

\subsubsection{X-ray Colors}
\begin{figure}
\centerline{\includegraphics[width=3in]{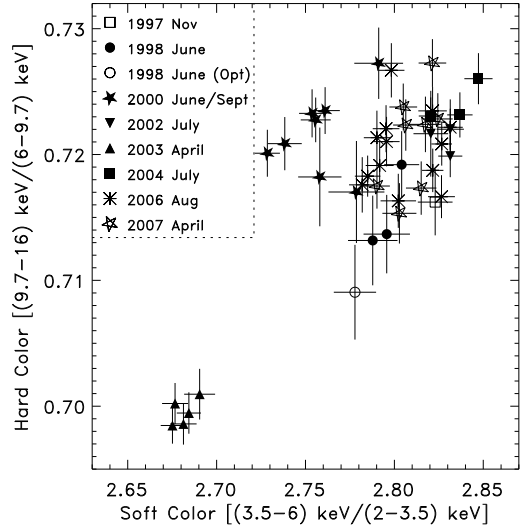}}
\caption{\footnotesize Average {\it RXTE}/PCA colors of \src~within 0.2 days of the times of the bursts shown in Fig. \ref{fluxdt}. \label{colors}}
\end{figure}

Figure \ref{colors} shows the average {\it RXTE}/PCA colors within 0.2 days of each burst. The individual 128 s colors (not shown here) show some scatter in the color-color diagram but all remain in the ``island'' state that typically characterizes bursting atoll LMXBs.  Evidently, the colors from 1997, 2002, 2004, 2006, and 2007 are similar. The colors from 1998 and 2000 are comparable, although the soft color from 2000 and the hard color from 1998 are slightly smaller than the colors from the other periods. On the other hand, the colors from the 2003 April are both significantly smaller, with fractional changes in the soft and hard colors of about 4\% and 3\%, respectively.  

\subsubsection{Broadband Spectra}
\label{bbs}
To test the extrapolation of our spectral model fits to energies below the
PCA and HEXTE passband, we also fitted the simultaneous 2002 July {\em
Chandra/RXTE} and 2003 April {\em XMM-Newton/RXTE} broadband spectra
(\S~\ref{chandrasect} \& \S~\ref{xmmsect}). In both cases, two
Comptonization components ({\tt comptt} in XSPEC; Titarchuk 1994)
were found necessary in order to accurately
model the data \citep{thompson05}. Each spectrum contains a component with
$\sim$0.1--0.2 keV seed photons, and a component with $\sim$0.8--0.9 keV
seed photons. One of these components is characterized by a $\sim$6 keV
electron plasma, and the other by a hotter electron plasma ($\sim$20 keV
or greater), to fit the hard tail. We could not unambiguously associate the high or low seed
photon temperatures with either the high or low plasma temperatures, however, due to the model degeneracy that is introduced by the use of two Comptonization components. 
Table 2 lists the best-fit parameter values for each spectrum and for the model coupling $kT_{\rm s,hot}$ to $kT_{\rm e,hot}$ and $kT_{\rm s,cold}$ to $kT_{\rm e,cold}$, plus fits to the 2003 April spectrum including an additional soft thermal component (see below).  

\begin{deluxetable*}{lcccc} 
\tablenum{2}
\tabletypesize{\scriptsize}
\tablecolumns{5}
\tablewidth{0pt}
\tablecaption{\sc{Broadband Spectral Fits to the Persistent Emission}} 
\tablehead{
\colhead{} \vline &
\colhead{2002 July} \vline &
\multicolumn{3}{c}{2003 April} \\
\colhead{Parameter} \vline  &
\colhead{2CTT\tablenotemark{a}} \vline &
\colhead{2CTT\tablenotemark{a}} &
\colhead{2CTT$+$DBB\tablenotemark{b}} & 
\colhead{2CTT$+$BB\tablenotemark{c}} 
}
\startdata                                          
$N_{\rm H}$ ($10^{21}/{\rm cm}^{2}$) & 3.33 $\pm$ 0.01  & 2.4 $\pm$ 0.1 & 3.73 $\pm$ 0.01 & 3.19 $\pm$ 0.01	\\
1. $kT_{\rm s}$ (keV)     & 0.21 $\pm$ 0.01  &  0.16 $\pm$ 0.01     & [0.19 $\pm$ 0.01]	& [0.15 $\pm$ 0.01]     \\
1. $kT_{\rm e}$ (keV)&  6.46 $\pm$ 0.03 & 5.9 $\pm$ 0.4 & 5.9 $\pm$ 0.3	& 6.0 $\pm$ 0.3		\\
1. $\tau$           & 4.72 $\pm$ 0.01 & 4.2 $\pm$ 0.1 & 4.4 $\pm$ 0.1 & 4.3 $\pm$ 0.1 \\
2. $kT_{\rm s}$ (keV) & 0.85$_{-0.01}^{+4.65}$  & 0.94 $\pm$ 0.02 & 0.75 $\pm$ 0.04 & 0.84 $\pm$ 0.05 \\
2. $kT_{\rm e}$ (keV)& $>29.5$   & 17.4$_{-1.8}^{+2.2}$ & 19.5$_{-1.8}^{+2.4}$ & 19.3$_{-2.1}^{+2.5}$           \\
2. $\tau$           & 1.65$_{-0.01}^{+8.93}$    & 2.9$_{-2.9}^{+0.3}$   & 2.9 $\pm$ 0.2	& 2.8 $\pm$ 0.2		\\
$\chi^{2}_{\nu}$ (d.o.f.) & 0.89 (1498)   & 1.22 (1152) & 1.09 (1151) & 1.02 (1151) \\
\enddata
\tablecomments{Errors correspond to the 90\% confidence interval for a single parameter. The 2002 July {\it Chandra/RXTE} spectrum had statistically significant data from 1--100 keV, and the 2003 April {\it XMM-Newton/RXTE} spectrum had statistically significant data from 0.5--100 keV. A multiplicative constant was included in the models to account for differences in the absolute flux normalizations between the instruments.}
\tablenotetext{a}{XSPEC model: {\tt tbabs*(comptt$_{1}$ + comptt$_{2}$)}. Each {\tt comptt} component assumes cylindrical geometry.}
\tablenotetext{b}{XSPEC model: {\tt tbabs*(comptt$_{1}$ + comptt$_{2}$ + diskbb)}. The temperature of the disk blackbody at the inner disk radius was tied to the cooler seed photon temperature (indicated by ``[\nodata]").}
\tablenotetext{c}{XSPEC model: {\tt tbabs*(comptt$_{1}$ + comptt$_{2}$ + bbody)}. The blackbody temperature was tied to the cooler seed photon temperature (indicated by ``[\nodata]").}
\end{deluxetable*}

\begin{figure}
\centerline{\includegraphics[width=3.3in]{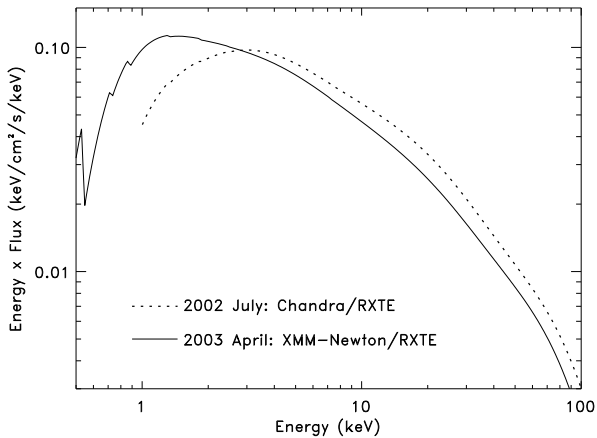}}
\caption{\footnotesize  Model spectra for the simultaneous {\em Chandra/RXTE} observations in 2002 July ({\em dotted} curve), and the simultaneous {\em XMM-Newton/RXTE} observations in 2003 April ({\em solid} curve). The {\em Chandra/RXTE} spectrum is restricted to $E > 1$ keV to reflect the energy range utilized in the fits; {\it XMM-Newton} provided good data down to 0.5 keV.\label{models}}
\end{figure}

The model spectra for both epochs are presented in Figure \ref{models}, which shows that the major model-independent differences are that the 2002 July spectrum peaks (in $\nu F_{\nu}$) at 1.3 keV, while the 2003 April spectrum peaks at 3.1 keV. {\it RXTE} observations without simultaneous low-energy coverage are, for the most part, insensitive to such changes in the soft flux, although the smaller soft colors (Fig. \ref{colors}) during 2003 April are suggestive of this difference. The model fitting the broadband 2003 April spectrum implies a 6\% increase in the flux relative to the fit with only {\it RXTE} data, illustrating how a wider passband can reveal additional flux not apparent in the model extrapolated from the 3--100 keV band. Still, the discrepancy between the persistent flux and the ``recurrence time flux'' is 30--40\% (Fig. \ref{resid}).

Relevant to the calculation of accurate bolometric fluxes for \src\ (see \S~\ref{esa}) is the question of additional soft spectral components. To test for the presence of such components, we separately tried adding a multicolor disk blackbody ({\tt diskbb} in XSPEC; Mitsuda et al. 1984) and a blackbody to our original spectral model, with the temperature of the soft component tied to the seed photon temperature of the cooler Comptonization spectral component. We interpret the model with a multicolor disk blackbody as an approximation to the accretion disk spectrum, or partial covering of the accretion disk due to the disappearance of the Comptonizing medium over a limited volume, and the model with a blackbody as partial covering of the neutron star surface, or an extended optically thick boundary layer. In this picture, the soft photons acting as seed photons therefore emerge without being Comptonized.

\begin{figure}
\centerline{\includegraphics[width=3.5in]{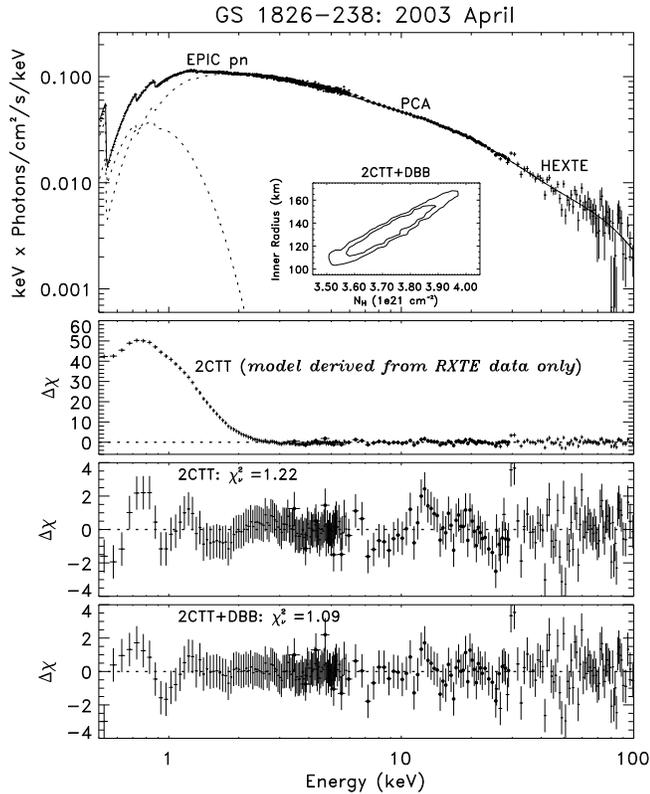}}
\caption{Unfolded 0.5--100 keV spectra and residuals for the 2003 April observations using the double Comptonization plus multicolor disk blackbody model (2CTT+DBB, {\it top and bottom panels}). The two middle panels show the fit residuals for the double Comptonization model derived using {\it RXTE} data alone ({\it upper middle panel}), and using both {\it XMM-Newton} and {\it RXTE} data ({\it lower middle panel}). The EPIC pn and HEXTE unfolded spectra are divided by 0.74 and 0.90, respectively, to account for the differences in the flux normalization between the instruments. The relatively large discrepancy between the PCA and EPIC normalizations has been noted previously; see e.g., \cite{yq03}. The residuals in the {\it XMM-Newton} data compared to the fit derived from {\it RXTE} data alone clearly indicate a substantial soft excess. By including {\it XMM-Newton} data, the double Comptonization model provides an acceptable fit, but the fits with an additional soft thermal component are superior. The 2CTT+BB residuals are not shown, although they are nearly indistinguishable from the 2CTT+DBB residuals. For clarity, only every tenth residual for the {\it XMM-Newton} data are shown. {\it Inset, top panel:} Confidence contours (68\% and 90\% levels) between the absorbing column density and the inner disk radius of the multicolor disk blackbody component (assuming a 6 kpc source distance and 60$^{\circ}$ inclination). \label{eufspec}}
\end{figure}

\begin{figure}
\centerline{\includegraphics[width=3.3in]{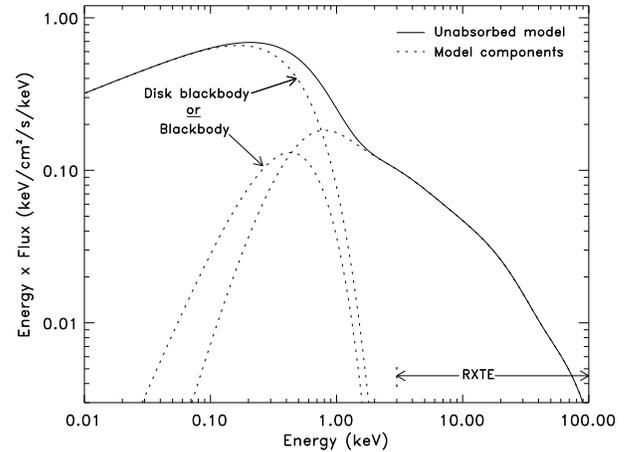}}
\caption{{\it Unabsorbed} 2003 April model spectrum containing an additional disk blackbody component {\it or} an additional blackbody component. The model is extrapolated to energies below the fitted range (0.5--100 keV). The {\em dotted} curves show the decomposition of the model components (the two Comptonized components are added together). The passband for {\em RXTE} is shown in the bottom of the figure, illustrating how a soft component could reside below {\it RXTE} coverage. Note how similar {\it absorbed} spectral shapes for $E>0.5$ keV are achieved with a model containing a combination of either a blackbody and an absorbing column density of $3.2 \times 10^{21}$ cm$^{-2}$, or a relatively brighter disk blackbody and a larger absorbing column density of $3.7 \times 10^{21}$ cm$^{-2}$. \label{diskbb}}
\end{figure}

For the 2002 July spectrum, we found that the additional component in each case (disk blackbody or blackbody) did not improve the fit to the data, yet allowed for a 4--5\% increase in the unabsorbed bolometric flux relative to the fit without the additional component. On the other hand, the fit to the 2003 April spectrum significantly improved in each case, with $F$-test probabilities of essentially zero that the improvement to the fit occurred by chance. The unfolded spectra and residuals for the double Comptonization plus multicolor disk blackbody model is shown in Figure \ref{eufspec} ({\it top} \& {\it bottom panels}). The residuals for the double Comptonization model derived using {\it RXTE} data alone, and using both {\it XMM-Newton} and {\it RXTE} data are also shown ({\it middle panels}). We stress, however, that the spectral fits for the models including a soft component should be considered {\it illustrative}. Accurately fitting a soft spectral component together with Comptonization components is problematic for larger seed photon temperatures because the {\tt comptt} model assumes a Wien input spectrum rather than a thermal spectrum. This approximation, while leading to negligible changes to the spectrum at high energies, leads to an underestimation of the low-energy flux.\footnote{Nevertheless, the Comptonization model is more appropriate than empirical models such as a cutoff power law which, for $\Gamma > 1$, diverges at low energies.} Compounding the issue is the interplay between the soft spectral component and the absorbing column density (see Fig. \ref{eufspec}, inset); although the single-parameter uncertainty in the absorbing column density is very small ($\sim$0.2\%), the two-parameter confidence contour between the absorbing column density and disk blackbody normalization ($\propto$$R_{\rm in}$) suggests a much larger uncertainty. Because reliable interpretation of the fits is not possible, we do not include the soft component normalization values in Table 2. Nevertheless, with the additional disk blackbody component the {\it unabsorbed} bolometric flux of the 2003 April spectrum is 3.7 $\times 10^{-9}$ erg cm$^{-2}$ s$^{-1}$, which is $\sim$50\% larger than the values obtained with only {\it RXTE}. With an additional blackbody, the unabsorbed flux is 3.3 $\times 10^{-9}$ erg cm$^{-2}$ s$^{-1}$, or $\sim$30\% larger than the values obtained with {\it RXTE} alone. Although the flux-recurrence time relation (Fig. \ref{fluxdt}) used the {\it absorbed} fluxes, a direct comparison can be made by noting that the {\it unabsorbed} fluxes for the other observing epochs are consistently $\sim$5\% larger (because a constant absorbing column density was used when extrapolating the 3--100 keV models to the wider 0.1--200 keV passband). The large fraction of soft flux contained in these components can clearly be seen in Figure \ref{diskbb}. Therefore, despite the limited ability to acquire physical understanding of a soft thermal component in the 2003 April spectrum, it is a distinct possibility that the presence of one can account for the apparent disparity between measured persistent flux and the flux level that is expected given the burst recurrence time (Fig. \ref{fluxdt}).

\subsection{Rapid Variability}
Each of the 18 average power spectra show very similar characteristic frequencies and rms amplitudes for the Lorentzian components representing the band-limited noise. The one major exception is the power spectrum from 2003 April, in which the characteristic frequencies are significantly higher. The 1998 June optical/$\rxte$~data show frequencies higher by a factor of $\sim$2. A single low frequency QPO is present in seven of the PSDs, and in all but two cases this QPO appears to be associated with $\nu_{h}$. Five of the PSDs have two low frequency QPOs, and in each case the higher (lower) frequency QPO also appears to be associated with $\nu_{h}$ ($\nu_{b}$). Similar low frequency QPOs have been observed in black holes and other neutron stars (e.g., Olive et al. 1998, Nowak 2000, Jonker et al. 2000), and even in previous observations of \src~\citep{barret00}. 

\begin{figure}
\centerline{\includegraphics[width=3in]{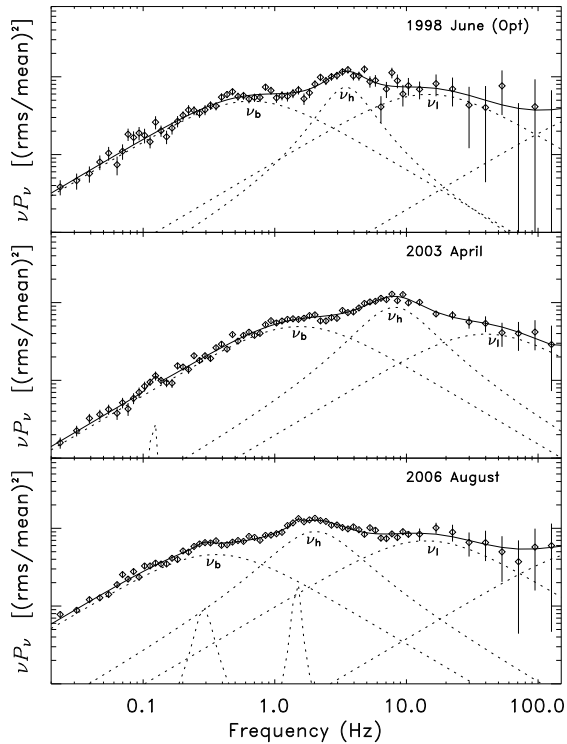}}
\caption{\footnotesize Example power spectra from 1998 June (during the time of the optical bursts), 2003 April, and 2006 August, with the individual Lorentzians superposed. The power spectra are presented in the $\nu P_{\nu}$ representation, which has the advantage of emphasizing higher frequency features and showing the maximum strength and frequency of the signal more directly. The components with characteristic frequencies of $\nu_{\rm b}$, $\nu_{\rm h}$, $\nu_{\rm l}$ are labeled (the high frequency feature centered on $\nu_{\rm u}$ is poorly constrained). The characteristic frequencies of the 2006 August PSD are more typical of the long-term average. The 2003 April PSD shows a dramatic increase in the characteristic frequencies, and the 1998 June PSD has moderately higher frequencies. \label{psdex}}
\end{figure}

\begin{figure}
\centerline{\includegraphics[width=3.5in]{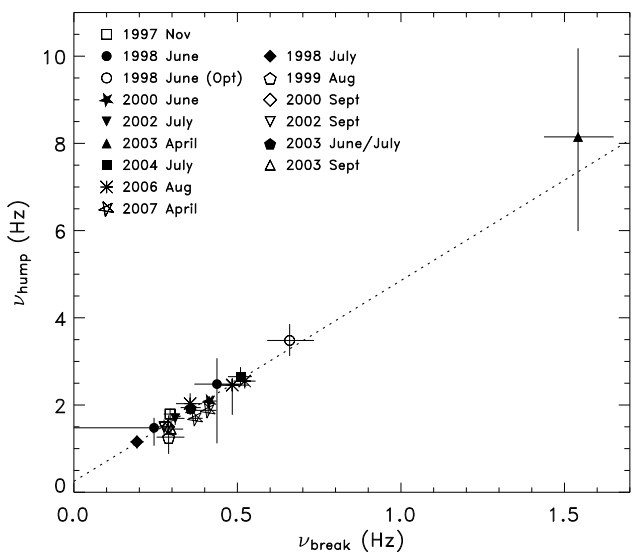}}
\caption{\footnotesize Characteristic frequencies of $\nu_{b}$ and $\nu_{h}$ for the 18 average {\it RXTE}/PCA PSDs. The {\em dotted} line illustrates the correlation among these parameters. The additional symbols on the right (compared to Figs. 2--4) are from observations lacking a recurrence time measurement.  \label{psdcorr}}
\end{figure}

Three of the power spectra are displayed in Figure \ref{psdex}. The top panel shows the power spectrum from 1998 June (optical), the middle panel from 2003 April, and the bottom panel from 2006 August. The latter is more indicative of the timing characteristics seen during the other {\it RXTE}\/ observation epochs. Figure \ref{psdcorr} shows the correlation between the two Lorentzian components that are present and well-constrained in all of the power spectra: $\nu_{b}$ and $\nu_{h}$. The dotted line shows the best-fit linear trend. Similar correlations between the Lorentzian characteristic frequencies has been seen in many other sources (e.g., Psaltis et al. 1999, Wijnands \& van der Klis 1999, Belloni et al. 2002, van Straaten et al. 2002), whether it be between the horizontal branch oscillation frequency and the lower frequency kHz QPO in Z-sources, or between the lower and upper frequency kHz QPOs in atolls, etc.  Given that these correlated timing features are rather common in neutron stars and black holes, the present correlation does not come as a surprise. 

\subsection{Early Ignition}
Although the X-ray spectral shape and timing characteristics are both
atypical during 2003 April (and, to a lesser extent, 1998 June), it is
also possible that some variation in the burst ignition properties
contributes to the unexpectedly short burst recurrence times measured
during those observations. \cite{bildsten00} pointed out that because the instability leading to bursts is a local phenomenon, the time for instability to develop depends on the accretion rate per unit area rather than the global accretion rate. A faster ignition time in 2003 April could thus be explained by accretion over a more limited area of the surface of the neutron star relative to the other observing epochs. Even if the instability begins sooner, we expect the burst will consume all of the fuel that has been accreted since the previous burst because, although the instability leading to the burst is a local phenomenon, the subsequent spreading burning front encompasses the entire stellar surface \citep{sp02,bs07}. In this case, we would expect the burst fluences to be reduced approximately to the ratio of the recurrence times, i.e., $\sim$3/4, assuming no change in the fuel composition at ignition, and a similar
accretion rate in 2003 April as for the observations at similar persistent
flux levels. Instead, the fluences for the 2003 April are roughly similar to the bursts at higher or lower accretion rates (Fig. \ref{fluxdt}, {\it middle} panel).


One way the burst fluences in 2003 April could still match the measured values for the other bursts despite $\sim$25\% less fuel is a change in the
composition of the fuel. The nuclear energy generation rate during bursts is $Q_{\rm nuc}=1.6+4X$, where $X$ is the mean hydrogen mass fraction in the
layer, and we assume $\approx$35\% energy loss due to neutrinos during
the $rp$-process burning (e.g., Fujimoto et al. 1987). A H-fraction that is
$\approx$35\% greater at ignition for the 2003 April bursts could thus 
compensate for the
smaller accreted mass. Naturally, the mean H-fraction will be higher 
for bursts with shorter recurrence times, since less time has passed for
steady H-burning to reduce the accreted composition. The time to burn all
the accreted hydrogen is $t_{\rm burn}=11 (Z/0.02)^{-1}(X_0/0.7)$~hr (e.g.,
Cumming \& Bildsten 2000), where $Z$ is the CNO mass fraction, and $X_0$ the
accreted composition. Assuming the accreted hydrogen begins burning
immediately upon being accreted, the mean H-fraction in the fuel layer at
ignition is then (for $\Delta t<t_{\rm burn}$): $X_0(1-0.5\Delta t/t_{\rm
burn})$. For $X_0=0.7$ and $Z=0.02$ this gives a H-fraction that is only
$\approx$6\% larger for the 2003 April bursts. Therefore, changes to the
composition of the accreted material cannot explain how the fluences of 
the bursts in 2003 April are in agreement with the other observing epochs.

\begin{deluxetable}{lccc} 
\footnotesize
\tablenum{3}
\tabletypesize{\scriptsize}
\tablecolumns{4}
\tablewidth{0pt}
\tablecaption{\sc{Burst Profile Comparison}} 
\tablehead{
\colhead{Parameter} &
\colhead{2003 April} &
\colhead{2000 June/Sept} &
\colhead{$>$2002 July\tablenotemark{a}}
}
\startdata   
$\tau$     & 39.7 $\pm$ 0.6 & 38.8 $\pm$ 1.2 & 40 $\pm$ 4 \\      
$\tau_{1}$ & 18.0 $\pm$ 0.9 & 17 $\pm$ 2 & 19 $\pm$ 2 \\  
$\tau_{2}$ & 47.3 $\pm$ 1.1 & 43.5 $\pm$ 0.8 & 45 $\pm$ 9 \\
$F_{\rm peak}$ & 24.5 $\pm$ 0.6 & 26.5 $\pm$ 0.7 & 25.5 $\pm$ 1.2 
\enddata
\tablecomments{The units of $\tau$, $\tau_1$, and $\tau_2$ are seconds, and the units of $F_{\rm peak}$ are $10^{-9}$ erg cm$^{-2}$ s$^{-1}$.}
\tablenotetext{a}{Excluding 2003 April.}
\end{deluxetable}

Significant changes to the {\it shape} of the burst light curves may indicate unusual or unexpected nuclear burning regimes. In order to characterize the burst profiles, we fitted a composite exponential curve to each burst, and we refer the reader to \cite{g06} for the details of the fitting procedure.  We compared the 2003 April burst decay parameters to those from the period of the most similar persistent flux (2000 June/September), and to the periods with the most similar burst recurrence times (all observations after 2002 July, except 2003 April).
This exercise revealed that the shapes of the burst profiles in 2003 April are qualitatively similar to the other epochs (Galloway et al. 2004; Heger et al. 2008). The parameter values are presented in Table 3. The only real indication of a significant difference between the light curves is in the second decay constant ($\tau_{2}$), which is somewhat longer for 2003 April than in 2000 June/September. However, these parameters do systematically depend on the recurrence time due to the changing hydrogen fraction (see Fig. 4 from Heger et al. 2008). A more appropriate comparison might be with the bursts at the closest recurrence time, i.e., the bursts with a recurrence time of $\sim$3.5 hr (2002 July and later, excluding 2003 April). Although this is a more inhomogeneous sample compared to 2003 April and 2000 June/September, $\tau_{2}$ for 2003 April is consistent with the others, though slightly higher (but not significantly) in the mean. There is a slightly longer second decay constant and lower peak flux for the 2003 April bursts compared to the bursts at the next nearest flux determined by {\it RXTE}, but, similar to the comparison of $\tau_{2}$, it is consistent with the bursts at the nearest recurrence time.



\section{Discussion}
During 2003 April, the recurrence times of thermonuclear bursts from \src~were unexpectedly short given the persistent flux measured by {\it RXTE} and the previously-determined monotonic relation between these two parameters. At the same time, the spectrum from a simultaneous {\it XMM-Newton} observation (\S~\ref{bbs}) suggests that an additional soft thermal component may be present at this time, which could increase the unabsorbed bolometric flux by up to 50\%. 
A significant piece of evidence supporting a redistribution of the accretion energy within the X-ray bands is the accompanying shift of the variability to faster timescales (Fig. \ref{psdcorr}), because correlations between higher power spectral frequencies and softer energy spectra are observed almost ubiquitously in black hole candidates and in many neutron stars. A direct relationship between QPO frequency and power-law index has been seen in many black hole candidates \cite[e.g.,][]{vig03,kal04,kal05,kal06,st06}, and in at least five neutron stars (e.g., Titarchuk \& Shaposhnikov 2005). Moreover, episodes of rare soft thermal components (interpreted as the accretion disk) lasting for several months in the black hole candidate GRS 1758--258 seem to be triggered by a decrease in the hard emission \citep{potts06}. Perhaps the most relevant example is from \cite{ford97}, however, who showed that there is a direct correlation between the flux of the blackbody component and the QPO frequency in the burster 4U 0614+091. Although high frequency QPOs are not observed in the \src~power spectra, the correlations between the frequencies of QPOs and broadband noise components in other sources make it tempting to speculate that the higher PSD frequencies in 2003 April are indeed coupled with a soft spectral component, the presence of which is presumably related to changes in the accretion geometry.

The correlations between faster variability and soft spectral components can be understood in the context of truncated disk models for accretion in X-ray binaries, which explain the low/hard to high/soft spectral transitions seen in both black hole and neutron star systems (for a recent review, see Done et al. 2008). For neutron star binaries at low $L/L_{\rm Edd}$ ($\sim$0.04--0.08 in \src, assuming $d=6$ kpc), such models consist of a thin accretion disk that is truncated at some radius far from the surface, probably due to evaporation of the accretion disk \citep{meyer00,mayer07}, and an optically thin\footnote{Note that the spectral parameters in Table 2 do not support an optically thin interpretation. We defer a detailed analysis of the spectrum using more sophisticated models such as {\tt eqpair} for a subsequent paper.} but geometrically thick hot inner flow that smoothly transitions to the boundary layer. Seed photons from the surface of the star and from the inner edge of the disk cool the inner flow. At higher accretion rates, the inner radius of the disk moves inwards, which reduces the volume of the region occupied by the hot inner flow. Assuming that most of the variability comes from this region, then a smaller volume would naturally account for the faster variability. In addition, the smaller disk radius will lead to a more prominent soft spectral component. 

One could argue that the soft photon flux should be present all the time, and perhaps just not detectable in the other observations due to poorer low-energy coverage. According to truncated disk models, however, at low accretion rates the accretion disk is expected to be far from the surface of the neutron star (e.g., Esin et al. 1997). If we assume that the power spectral break frequency $\nu_{\rm b}$ is associated with the inner disk radius and the orbital frequencies are Keplerian ($\nu_{\rm Kep} \propto R^{-3/2}$), we expect the inner disk radius in 2002 July, for example, to be $\approx$3 times larger than in 2003 April, and the integrated disk flux to be $\approx$3 times smaller (Frank et al. 1992, p.73). This is roughly consistent with the ratio of the absorbed soft fluxes shown in Fig. \ref{models}.

Given that the 2003 April power spectral frequencies are the highest, and at the same time the burst recurrence time suggests the accretion rate is the highest, it is worth considering that an alternative indicator for the mass accretion rate may be the timing data rather than the X-ray flux. For example, it has been proposed that $\nu_{\rm b}$ is positively correlated with the mass accretion rate (e.g., van der Klis 1994). If this were the case, however, we would expect a monotonic relationship between $\nu_{\rm b}$ and burst recurrence time (which we also assume to trace mass accretion rate). Such a relationship does not exist. To see this, one must only consider the data from 1998 June (optical); the break frequency is the second highest value even though the inferred accretion rate is the third lowest (only 1997 November and 1998 June have longer burst recurrence times, Fig. \ref{fluxdt}). On the other hand, the two observing epochs showing the largest disparity between the recurrence time flux and the observed flux (Fig. \ref{resid}), i.e., 1998 June (optical) and 2003 April, also exhibit the highest break frequencies. Therefore, it may be the case that $\nu_{\rm b}$ is a better tracer of the soft thermal flux and not the bolometric flux, although the absence of low-energy spectral coverage in the majority of \src~observing epochs precludes a detailed comparison.

An alternative explanation is that the bursts are igniting earlier, perhaps because the area over which the accretion occurs has decreased; however, any plausible mechanism to ignite the bursts earlier would also give rise to bursts that were significantly less energetic than others at comparable X-ray fluxes, whereas we find bursts in 2003 April with fluences comparable to those in other epochs. Thus, we conclude that the X-ray flux measured by {\it RXTE} in 2003 April underestimates the bolometric flux to a much greater degree than in other epochs, likely due to the presence of enhanced X-ray emission below $\sim$2 keV. This implies that although the X-ray flux measured by {\it RXTE} in 2003 April was similar to what was observed in 2000 June/September, the accretion rate (and ``true'' bolometric flux) in 2003 April must have been at its highest level because the recurrence time was the smallest. This idea is consistent with truncated disk models, which explain the emergence of the soft spectral component by a reduction of the inner disk radius at high accretion rates. In this scenario, the previous and subsequent {\it RXTE} observations of \src~fortuitously maintained a fairly strict correspondence between X-ray flux in the {\it RXTE} passband and accretion rate due to a relatively constant spectral shape. Our results indicate that the spectral shape in \src~starts to undergo a significant transformation when the bolometric flux is above about $3.5 \times 10^{-9}$ erg cm$^{-2}$ s$^{-1}$, or when the luminosity is above $1.5 \times 10^{37}$ erg s$^{-1}$ $(d/6~{\rm kpc})^{2}$.


There are other factors that could play a role during the periods of discrepant bursting behavior. One possibility is that the photon emission from \src~may be anisotropic. Such emission is expected any time the scattering optical depth is much lower over a limited range of solid angles than in other directions \citep{king01}, and includes any source with an accretion disk. Recent work by Heger et al. (2008) suggests that emission anisotropy does indeed influence the observational characteristics of \src. In order for the burst timing and energetics from {\it RXTE} observations between 1997 November to 2002 July to match theoretical ignition models, they found that $\xi_{\rm p}/\xi_{\rm b} =1.55$. Different anisotropy factors for the burst and persistent emission are expected because the radiation can be attributed to geometrically distinct regions \citep{fuji88}. If we assume that the burst emission is isotropic, the results of Heger et al. (2008) indicate that the persistent luminosity may be underestimated by a factor of 1.55 in all observing epochs. To determine if anisotropic emission affected the 2003 April data to a higher degree than the other time periods, however, would require the results to be extended to include this epoch.

\acknowledgements
We thank Tod Strohmayer for assisting with determining the correction factor to apply to the observations with offset pointing, and Simone Migliari for providing the colors of the Crab that were used to correct for PCA gain changes. We also appreciate significant contributions from J\"{o}rn Wilms and Katja Pottschmidt. This work was supported in part by NASA contract NAS5-30720.

\end{document}